\documentclass[aps,pra,twocolumn]{revtex4-2}  
\usepackage{orcidlink}
\usepackage{amsmath}
\usepackage{mathrsfs}
\usepackage{amssymb}
\usepackage[utf8]{inputenc}
\usepackage{graphicx,epstopdf,epsfig}
\usepackage{bm}
\usepackage{color}
\usepackage{times}
\usepackage{hyperref,url}
\usepackage{xcolor,soul}
\usepackage{xspace}
\usepackage{changepage}
\usepackage{lipsum}
\usepackage{comment}
\usepackage{soul} 
\usepackage{textcomp}

\newcommand{\bra}[1]{{\langle #1 \vert}} 		
\newcommand{\ket}[1]{{\vert #1 \rangle}} 		
\newcommand{\braket}[1]{\langle #1 \rangle}     

\renewcommand{\eqref}[1]{Eq.\ \ref{#1}}

\newcommand{\Forster}{F\"orster\xspace}        %

\begin{document}

\title{Interspecies \Forster resonances for Rb-Cs Rydberg $\mathrm{d}$-states for enhanced multi-qubit gate fidelities}
\author{P. M. Ireland\,\orcidlink{0009-0008-9455-9134}}
\author{D. M. Walker\,\orcidlink{0009-0009-0100-6555}}
\author{J. D. Pritchard\,\orcidlink{0000-0003-2172-7340}}
\email{jonathan.pritchard@strath.ac.uk}
\affiliation{Department of Physics, SUPA, Strathclyde University, Glasgow, G4 0NG, UK}

\begin{abstract}
We present an analysis of interspecies interactions between Rydberg $d$-states of rubidium and cesium. We identify the F\"orster resonance channels offering the strongest interspecies couplings, demonstrating the viability for performing high-fidelity two- and multi-qubit $C_kZ$ gates up to $k=4$, including accounting for blockade errors evaluated via numerical diagonalization of the pair-potentials. Our results show $d$-state orbitals offer enhanced suppression of intraspecies couplings compared to $s$-states, making them well suited for use in large-scale neutral atom quantum processors.
\end{abstract}

\maketitle

\section{Introduction}

Neutral atom arrays provide a versatile platform for performing both programmable quantum simulation and quantum computation \cite{saffman10,adams19,browaeys20,morgado21}. Arrays of optical tweezers offer a scalable route to creating deterministically loaded, defect-free qubit registers in up to three dimensions \cite{endres16,barredo16,barredo18,kumar18}. Interactions between qubits can be engineered using highly-excited Rydberg states to perform high-fidelity two- and multi-qubit gate operations \cite{isenhower10,maller15,zeng17,graham19,levine19,fu22,mcdonnell22,pelegri22} for implementing digital quantum algorithms \cite{graham22,bluvstein22}, or for encoding solutions to classical optimization problems using coherent quantum annealing \cite{pichler18,ebadi22,nguyen23}. Using this architecture for quantum simulation \cite{ebadi21,scholl21} has enabled novel topological phases \cite{leseleuc19a} and quantum spin liquids \cite{semeghini21} to be observed. Combining these approaches with dynamically reconfigurable tweezers \cite{bluvstein22} provides a route to observation of fast-scrambling dynamics \cite{hashizume21} or efficient implementation of low-density parity check codes for quantum error correction \cite{xu23}, with recent demonstrations of transversal gates between logical qubits making a first step towards fault-tolerant operation \cite{bluvstein24}.

Whilst there has been significant progress to advance two-qubit gate fidelities $\mathcal{F}>99.5\%$ \cite{evered23,ma23} along with recent demonstrations of scalable non-destructive readout \cite{nikolov23} and mid-circuit measurement \cite{deist22, graham23, lis23}, most experiments have used only a single atomic species. This introduces limitations in localized qubit readout due to cross-talk between atoms, and in multi-qubit gate operations where fidelities are limited by parasitic interactions between Rydberg states\cite{levine19,ebadi22,pelegri22}. 

\begin{figure}[t!]
\includegraphics{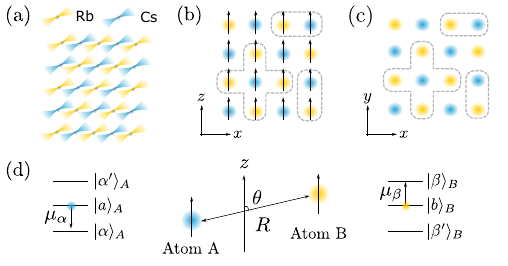}
\caption{Dual-species interactions. (a) Schematic of a dual species array of Cs and Rb atoms using state-selective optical tweezers. For performing interspecies $CZ$ or $C_4Z$ gates (indicated by dashed outlines), the orientation of the internuclear axis relative to the quantization axis ($\hat{z}$) is important, with quantization along the tweezer axis in (b) leading to interactions at a range of angles, whilst quantization perpendicular to the array plane (c) ensures all interactions occur for $\theta=90^\circ$. (d) Interaction angle $\theta$ relative to the quantization axis, with an example \Forster{}-resonant state occuring when the initial pair state $\ket{a}_A\ket{b}_B$ is dipole-coupled to $\ket{\alpha}_A\ket{\beta}_B$ with a small energy defect $\Delta$. }
\label{fig:schematic}
\end{figure}

One approach to overcome this limitation is to use dual-species arrays, which naturally provides a separation in readout wavelengths to suppress cross-talk, whilst allowing engineering of different inter and intra-species couplings. Additionally, this offers a route to universal quantum computation using globally driven pulses \cite{cesa23}. Early work demonstrated two-qubit gates and arrays between different isotopes of rubidium (Rb) \cite{zeng17,sheng22}, with recent developments showing continuous loading and measurement-feedback onto a cesium (Cs)-Rb array \cite{singh22,singh23}. This same approach of using heterogeneous tweezers has enabled assembly of polar molecules \cite{liu18,liu19,cairncross21,yu21,ruttley23}, leading to first demonstrations of hybrid systems \cite{zhang22,wang22} based on coupling a Rydberg atom to a polar molecule \cite{guttridge23}.

Previous theoretical work exploring dual-species Rydberg interactions has focused on studying Rydberg interactions in $s$-state orbitals for Rb-Cs \cite{beterov15} and $s$ and $d$ states for Rb-K atoms \cite{samboy17,otto20}, where suitable F\"orster resonances have been identified to obtain strong intraspecies couplings with dispersion coefficients calculated in the asymptotic limit. This forms the basis of proposals for quantum non-demolition readout \cite{beterov15}, quantum error correction using dual-species surface codes \cite{auger17}, and high-fidelity heteronuclear C$_2$NOT$^2$ gates \cite{farouk23} which suppress the unwanted interspecies target-target and control-control interactions.

In this paper we extend the dual-species analysis to study interspecies interactions between Cs and Rb Rydberg atoms in the $d$-state orbitals, utilising open source libraries \cite{weber17,sibalic17} for calculating Rydberg-atom interaction potentials via direct diagonalization of the dipole-dipole Hamiltonian to go beyond evaluating the asymptotic dispersion coefficients. We identify suitable F\"orster resonances for maximising interspecies coupling strengths, carefully considering the angular dependence to address applications of performing two- and multi-qubit gates in neutral atom arrays for different quantization axis choices as illustrated in Fig.~\ref{fig:schematic}. We quantify blockade leakage errors for different \Forster-pair states, and further evaluate the intrinsic gate errors based on the canonical three-pulse controlled phase gate protocol \cite{jaksch00} for $C_kZ$ gates up to $k=4$, demonstrating enhanced fidelity for multi-qubit operations compared to single-species approaches. We show the advantages of using $d$-state rather than $s$-state resonances identified in previous studies due to the inherent reduction in intraspecies couplings. These results are relevant to teams working in both quantum computation and simulation, and provide a guide to appropriate state choice and polarization when designing dual-species experiments.

The paper is organized as follows. In Sec.~\ref{sec:Vdd} we provide a brief introduction to Rydberg atom interactions, then in Sec.~\ref{sec:dres} we identify the dominant  F\"orster resonances for the Rb-Cs $d$-states, studying the angular dependence and comparing to previously identified $s$-state resonances.  In Sec.~\ref{sec:leakage} we evaluate the blockade and leakage errors associated with \Forster-resonant pairs, considering the performance for achieving both strong blockaded interactions for Rb-Cs pairs and weak intraspecies Rb-Rb and Cs-Cs couplings. From this analysis the Rb $59d_{5/2}$ - Cs $6d_{3/2}$ resonance offers the best performance for extension to $C_kZ$ gates and in Sec.~\ref{sec:gate} we provide estimates of gate fidelity, before summarizing our findings in Sec.~\ref{sec:conclusion}.

\section{Rydberg Atom Interactions}
\label{sec:Vdd}
To begin we briefly summarize the methods used to calculate Rydberg atom interactions. For a pair of atoms separated by distance $R$ at angle $\theta$ with respect to the quantization axis, as illustrated in Fig.~\ref{fig:schematic}(b), the dipole-dipole interaction is given by \cite{reinhard07}
\begin{equation}\label{eq:dipole_dipole}
    V(R) = \frac{ \bm{\mu}_\text{A} \cdot \bm{\mu}_\text{B} - 3 (\hat{\bm{n}}\cdot\bm \mu_\text{A} ) (\hat{\bm{n}}\cdot\bm \mu_\text{B} )}{R^3},
\end{equation}
where $\bm\mu_{A}$ and $\bm{\mu}_{B}$ are the electric dipole moments for atom A and atom B respectively for coupling target pair states $\vert a\rangle_A\vert b\rangle_B$ to pair states $\vert \alpha\rangle_A\vert \beta\rangle_B$, and $\hat{\bm n}$ is the unit vector along the atom separation axis $R$.

The sign and strength of the resulting interaction is determined by the dipole coupled pair state $\vert{\alpha^A\beta^B}\rangle\equiv\vert \alpha\rangle_A\vert \beta\rangle_B$ with the smallest energy defect $\Delta = (E_\alpha-E_a)+(E_\beta-E_b)$ with respect to the target pair state $\ket{ab}$. For $|\Delta|\gg V(R)$ the resulting interaction is in the van der Waals regime of the form with energy shifts scaling as $C_6/R^6$, whilst for $|\Delta|<V(R)$ the system experiences a strong first-order energy shift into branches with energies of $\pm C_3/R^3$. Alternatively, dc electric fields can be used to tune a near-resonant pair to achieve $\Delta=0$ \cite{reinhard08,ryabtsev10,ravets14}.

To estimate the sign and magnitude of the coupling strengths for a given pair state, it is instructive to calculate the $C_{3,k}$ channel coefficient introduced in Ref.~\cite{beterov15} defined as

\begin{equation}
C_{3,k}=\frac{e^2}{4\pi\epsilon_0}\frac{\langle \alpha_k\vert\vert r_A\vert\vert a\rangle\langle\beta_k\vert\vert r_B \vert\vert b\rangle}{\sqrt{(2j_{\alpha_k}+1)(2j_{\beta_k}+1)}},
\end{equation}
where $\ket{\alpha^A_k\beta^B_k}$ is a Rydberg pair state which is dipole coupled to the target pair state $\ket{a^Ab^B}$, and $\langle \phi_k\vert\vert r_X\vert\vert \psi\rangle$ is the reduced matrix element for transitions between states $\psi$ and $\phi_k$ in the fine-structure basis for atom $X$ associated with interaction channel $k$. $j_\psi$ is the total angular momentum quantum number associated to the state $\psi$.

Asymptotic dispersion coefficients $C_6$ and $C_3$ can be calculated by summing over the relevant dipole-coupled pair states \cite{singer05,beterov15}, however for the typical atomic separations used in current tweezer array experiments $3<R<15~\mu$m accurate calculation of the resulting interatomic interactions requires exact diagonalization of the resulting pair-state Hamiltonian including higher-order dipole-quadrupole and quadrupole-quadrupole terms. Below we use the \emph{pair-interaction} library \cite{weber17} to evaluate Rydberg state couplings, which exploits measured quantum defects and numerical integration of the atomic wavefunctions based on model potentials to calculate Rydberg state energies and matrix elements respectively. 

To engineer strong interspecies couplings it is necessary to identify \Forster{} resonant pair states that have small energy defects $\Delta$ and large coupling strenthgs $C_{3,k}$ \cite{beterov15}. For two independent species or isotopes, this is achieved by choosing relevant quantum numbers including principal quantum number and orbital angular momentum, resulting in a large degree of tunabiltiy and control in choosing Forster resonant pairs.  
For the intraspecies couplings however, the energy defects of a given pair is determined entirely by the inherent quantum defects for each species, meaning there is typically only a single value of $n$ which is near-resonant and that predominantly the intraspecies interactions are in the weaker van der Waals regime which drops off rapidly with $R$.

\section{Rubidium-Cesium $d$-state F\"orster Resonances}\label{sec:dres}

We identify suitable \Forster-resonant pairs of Rb and Cs in $d$ orbital angular momentum states by calculating the energy defect $\Delta$ and channel coefficient $C_{3,k}$ for all dipole-coupled pair states from atoms initially in $d_{3/2}$ or $d_{5/2}$. Due to the large number of possible \Forster resonant pair states, we follow the approach of Ref.~\cite{beterov15} to consider only resonances with $50\le n \le 90$ that have $\vert C_{3,k}\vert>1$~GHz$\,\mu$m$^3$ and an energy defect $\vert \Delta\vert <0.005$ of the level spacing. Complete tables of resonances are given in the Appendix, but we focus here on identifying the dominant channels for different combinations of values of the total angular momentum quantum number $j$.

\subsection{Rb $\mathrm{d}_{5/2}$ - Cs $\mathrm{d}_{5/2}$}
The $n_\mathrm{Rb}\mathrm{d}_{5/2}n_\mathrm{Cs}\mathrm{d}_{5/2}$ pair-states are dipole coupled through a total of 9 possible channels associated with angular quantum numbers of $n'_\mathrm{Rb}p_{3/2}n'_\mathrm{Cs}\mathrm{p}_{3/2}$, $n'_\mathrm{Rb}\mathrm{f}n'_\mathrm{Cs}\mathrm{p}_{3/2}$, $n'_\mathrm{Rb}\mathrm{p}_{3/2}n'_\mathrm{Cs}\mathrm{f}$ and $n'_\mathrm{Rb}\mathrm{f}n'_\mathrm{Cs}\mathrm{f}$ where the $\mathrm{f}$-states can be either the $j=5/2$ or $j=7/2$ magnetic sublevels. 
From the resonances listed in Tab.~\ref{tab:DD} there is only a single resonance via the $n'_\mathrm{Rb}\mathrm{p}_{3/2}n'_\mathrm{Cs}\mathrm{p}_{3/2}$ channel, whilst the strongest coupling coefficient's $C_{3,k}$ occur for the $(n_\mathrm{Rb}+1)\mathrm{p}_{3/2}(n_\mathrm{Cs}-2)\mathrm{f}_{7/2}$ channel. Additional dominant channels are $(n_\mathrm{Rb}-2)\mathrm{f}(n_\mathrm{Cs}-2)\mathrm{f}$ and $(n_\mathrm{Rb}-2)\mathrm{f}(n_\mathrm{Cs}+2)\mathrm{p}_{3/2}$.

\begin{figure}[t]
\includegraphics[width=0.5\textwidth]{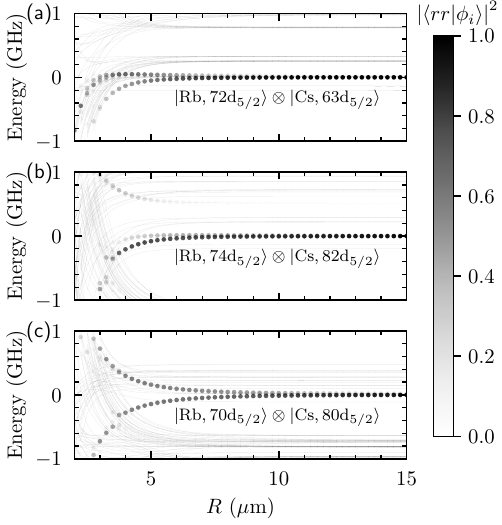}
\caption{Example Rb-Cs $n_\mathrm{Rb}\mathrm{d}_{5/2}n_\mathrm{Cs}\mathrm{d}_{5/2}$ resonances. (a) Rb $72\mathrm{d}_{5/2}$-Cs $63\mathrm{d}_{5/2}$, resonant with Rb $70\mathrm{f}$-Cs $61\mathrm{f}$ (b) Rb $74\mathrm{d}_{5/2}$-Cs $82\mathrm{d}_{5/2}$, resonant with Rb $72\mathrm{f}$-Cs $84\mathrm{p}_{3/2}$ (c) Rb $70\mathrm{d}_{5/2}$-Cs $80\mathrm{d}_{5/2}$, resonant with Rb $71\mathrm{p}_{3/2}$-Cs $78\mathrm{f}$. Calculated for $\theta=0^\circ$ and $m_j^\mathrm{Rb}=m_j^\mathrm{Cs}=+5/2$.}
\label{fig:DD}
\end{figure}

Using these states we now examine the characteristic pair-potentials for each of these channels via direct matrix diagonalization, calculating potentials for $\theta=0^\circ$ and considering atoms in magnetic sublevels $m_j^\mathrm{Rb}=m_j^\mathrm{Cs}=+5/2$. For the $(n_\mathrm{Rb}-2)\mathrm{f}(n_\mathrm{Cs}-2)\mathrm{f}$ channel, Fig.~\ref{fig:DD}(a) shows the coupling for Rb $72\mathrm{d}_{5/2}$-Cs $63\mathrm{d}_{5/2}$ which is resonant with Rb $70\mathrm{f}$-Cs $61\mathrm{f}$ states with $\Delta/2\pi=19.1$~MHz, with datapoint intensity coloured to reflect the overlap of the target state with the calculated eigenstate $\ket{\phi_i}$ as $\vert\bra{rr}\phi_i\rangle\vert^2$. Whilst this resonance possesses a strong channel coefficient of $\vert C_{3,k}\vert=7.64~$GHz$\,$\textmu m$^3$, it is evident that rather than a strong resonant interaction at short range ($R<10\,$\textmu m), the off-resonant coupling to other pair-states suppresses the shift of the upper excitation branch, leading to a weak interaction that is poorly suited to realising strongly blockaded excitation.

In Fig.~\ref{fig:DD}(b) we show the Rb $74d_{5/2}$-Cs $82d_{5/2}$ state, which is near resonant with Rb $72f$-Cs $84p_{3/2}$ states. Around $R=4-6~\mu$m we observe the $f$ states split due to coupling with other branches, resulting in a strongly-shifted upper excitation branch and mixing in a third eigenstate that is only weakly shifted and preventing use of the $(n_\mathrm{Rb}-2)f(n_\mathrm{Cs}+2)p_{3/2}$ resonances for high-fidelity gates.  

In Fig.~\ref{fig:DD}(c) we show the $72d_{5/2}$-Cs $80d_{5/2}$, which is strongly coupled to the resonant Rb $72f$-Cs $84p_{3/2}$ states with $\vert C_{3,k}\vert = 23~\mathrm{GHz}\,\mu\mathrm{m}^3$. This reveals the expected symmetric splitting and $1/R^3$ scaling, making the $(n_\mathrm{Rb}+1)p_{3/2}(n_\mathrm{Cs}-2)f_{7/2}$ resonances ideal for engineering strong interspecies couplings at $\theta=0^\circ$. These results highlight the importance not only of evaluating the pair defects and channel couplings, but also examining the real pair curves over the desired separation ranges.

\begin{figure}[t]
\includegraphics[width=0.5\textwidth]{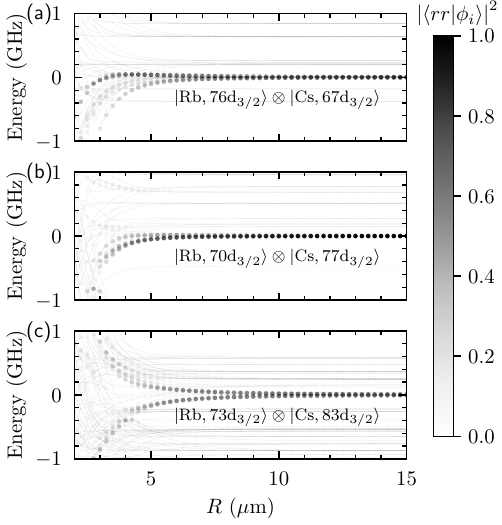}
\caption{Example Rb $d_{3/2}$-Cs $d_{3/2}$ \Forster resonances. (a) Rb $76d_{3/2}$-Cs $67d_{3/2}$, resonant with Rb $74f_{5/2}$-Cs $65f_{5/2}$ (b) Rb $70d_{3/2}$-Cs $77d_{3/2}$, resonant with Rb $68f_{5/2}$-Cs $79p_{1/2}$ (c) Rb $73d_{3/2}$-Cs $83d_{3/2}$, resonant with Rb $74p_{1/2}$-Cs $81f_{5/2}$. Calculated for $\theta=0^\circ$ and $m_j^\mathrm{Rb}=m_j^\mathrm{Cs}=+3/2$.}
\label{fig:dd}
\end{figure}

\subsection{Rb $d_{3/2}$ - Cs $d_{3/2}$}
For the Rb $d_{3/2}$-Cs $d_{3/2}$ states there are again 9 possible channels associated with orbital angular momentum quantum numbers of $n'_\mathrm{Rb}pn'_\mathrm{Cs}p$, $n'_\mathrm{Rb}pn'_\mathrm{Cs}f_{5/2}$, $n'_\mathrm{Rb}f_{5/2}n'_\mathrm{Cs}p$ and $n'_\mathrm{Rb}f_{5/2}n'_\mathrm{Cs}f_{5/2}$ where $p$ can be either $j=1/2,3/2$. Table~\ref{tab:dd} shows the filtered resonances, with the strongest $\vert C_{3,k}\vert$ channel coupling via the $(n_\mathrm{Rb}+1)p_{1/2}(n_\mathrm{Cs}-2)f_{5/2}$ resonances.

In Fig.~\ref{fig:dd} example resonances for the different channels are plotted, with similar behavior observed as for the $d_{5/2}-d_{5/2}$ resonances where the channels with pair-states resonant with Rb $f_{5/2}$ states (shown in (a) and (b)) provide unsuitable resonances at short range. The strongest resonance is for Rb $73d_{3/2}$-Cs $83d_{3/2}$ which is coupled to Rb $74p_{1/2}$-Cs $81f_{5/2}$ with $\Delta/2\pi=-0.84$~MHz and $C_{3,k}=-27.3~$GHz$\,\mu$m$^3$ plotted in Fig.~\ref{fig:dd}(c) showing a strong \Forster{} resonant coupling at short range, with similar behavior observed for resonances for pairs involving Rb $p_{3/2}$ state but with a weaker coupling due to the larger defects. At short range ($R<5\mu$m), comparing the result here to the approximately equivalent resonance shown in Fig.~\ref{fig:DD}(c) at comparable quantum numbers, the $d_{3/2}-d_{3/2}$ resonance mixes with more pair states than the $d_{5/2}-d_{5/2}$ case, leading to the two characteristic branches splitting with the target pair state now appearing across a larger number of pair eigenstates.

\subsection{Rb $d_{3/2}$ - Cs $d_{5/2}$ and Rb $d_{5/2}$ - Cs $d_{3/2}$}
\Forster resonances for the Rb $d_{3/2}$-Cs $d_{5/2}$ and Rb $d_{5/2}$-Cs $d_{3/2}$ states are presented in Tab.~\ref{tab:dD} and ~\ref{tab:Dd} respectively. These mixed $j$ pairs follow a similar trend to those above, with the strongest resonances observed for $(n_\mathrm{Rb}+1)p_{j_\mathrm{Rb}-1}(n_\mathrm{Cs}-2)f_{j_\mathrm{Cs}+1}$ couplings, and resonances with Rb $p$ and Cs $f$ states producing the desired $1/R^3$ resonant splitting at $\theta=0^\circ$. In comparison to the Rb $d_{3/2}$-Cs $d_{5/2}$ states, Rb $d_{5/2}$-Cs $d_{3/2}$ pairs typically have larger $\Delta$ values, however there are more states with strong $C_{3,k}$ at lower values of $n_\mathrm{Rb}$.

As will be shown below in Sec.~\ref{sec:gate}, the resonances via Rb $d_{3/2}$ states are also less favorable for scaling to multi-qubit interactions due to the presence of a near \Forster resonance for Rb $58d_{3/2}$-Rb $58d_{3/2}$, leading to anomalously large intraspecies $C_6$ coefficients for nearby $n_\mathrm{Rb}d_{3/2}$ states.

\subsection{Angular Dependence}
In the analysis above we consider atoms at $\theta=0^\circ$, which results in the total magnetic quantum number $M=m_j^A+m_j^B$ being preserved by the dipole-dipole interaction with $\Delta M=0$ selection rules. For experiments in 2D tweezer arrays, this condition is only met for the case of atoms oriented with the internuclear axis along the quantization axis. For performing multi-qubit gate operations, or exploiting long-range couplings for quantum simulation, placing a quantization axis parallel to one of the axis coordinates results in atoms interacting across a range of angles from $\theta=0^\circ$--$90^\circ$ as shown in Fig.~\ref{fig:schematic}(b). Alternatively, the quantization axis can be aligned normal to the array plane, meaning all atoms interact at $90^\circ$ as shown in Fig.~\ref{fig:schematic}(c), providing an isotropic coupling of an atom to its neighbors, where at this angle $V(R)$ couples pair states with $\Delta M=0,\pm2$. In this section we explore the angular dependence of the $d$-state \Forster{} resonances, considering the impact on choice of magnetic sublevel.

\begin{figure}[t!]
\includegraphics[width=\linewidth]{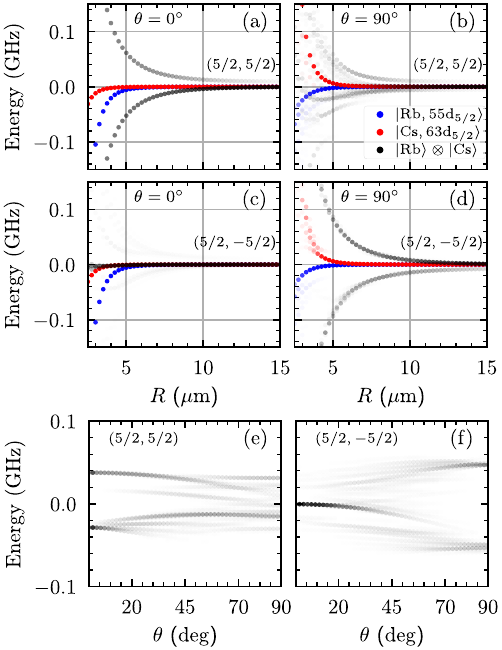}
\caption{Angular dependence of the Rb $55d_{5/2}$ - Cs $63d_{5/2}$ \Forster resonance for  different $(m_j^\mathrm{Rb},m_j^\mathrm{Cs})$. (a) $\theta=0^\circ$, $(5/2,3/2)$ (b) $\theta=90^\circ$, $(5/2,5/2)$ (c) $\theta=0^\circ$, $(5/2,-5/2)$ (d) $\theta=90^\circ$, $(5/2,-5/2)$. Interspecies pair-potentials are plotted black with opacity proportional to state overlap. Also shown are intraspecies interaction curves are shown for Rb-Rb in $55d_{5/2}$ (red) and Cs-Cs $63d_{5/2}$ (blue). (e,f) The variation of energy eigenstates with $\theta$ for $(5/2,\pm5/2)$ calculated at $R=6~\mu$m revealing the \Forster resonance can only be used at $0^\circ$ and $90^\circ$.}
\label{fig:potential_d_d}
\end{figure}

Figure~\ref{fig:potential_d_d} shows angular potentials for the \Forster resonance between Rb $55d_{5/2}$ - Cs $63d_{5/2}$ for different combinations of $(m_j^\mathrm{Rb},m_j^\mathrm{Cs})=(5/2,\pm5/2)$ and for angles $\theta=0^\circ$ and $90^\circ$. Considering first the $(5/2,5/2)$ combination, Fig.~\ref{fig:potential_d_d}(a) shows at $0^\circ$ we obtain a strong \Forster resonance splitting, however at $90^\circ$ the additional coupling terms driving $\Delta M =\pm2$ cause the target Rydberg states to be mixed with a large number of other pair eigenstates, several of which have weak or vanishing interaction shifts. Conversely, for the $(5/2,-5/2)$ combination with $M=0$ we see a strong suppression of the interaction at $0^\circ$ in Fig.~\ref{fig:potential_d_d}(c) leading to a flat potential curve for $R>5~\mu$m due to the presence of a \Forster-zero in the interaction channel \cite{walker05}, whilst for $90^\circ$ we recover the desired \Forster splitting as shown in Fig.~\ref{fig:potential_d_d}(d). Comparing the results in (a) and (d) shows that the $\theta=90^\circ$ offers improved symmetry between upper and lower excitation branches of the pair potentials with respect to the unshifted pair state, extending the useful blockade range which is limited by the smallest shift of either branch. This shows that operation at $90^\circ$ with a $(5/2,-5/2)$ combination is preferable not only for achieving isotropic couplings across the array, but also for improved long-range interactions.

Alongside the Rb-Cs pair potentials, the figures also show intraspecies interaction curves for Rb $55d_{5/2}$ - $55d_{5/2}$ in blue and Cs $63d_{5/2}$-$63d_{5/2}$ in red. For both Fig.~\ref{fig:potential_d_d}(a) and (d) with strong interspecies \Forster resonance, we see the intraspecies interaction curves remain flat until $R\lesssim5~\mu$m before showing a van der Waals like $C_6/R^6$ shift. Despite having a lower quantum number of $n_\mathrm{Rb}=55$, the Rb-Rb $C_6$ coefficient is comparable to that of the $n_\mathrm{Cs}=63$ due to the Rb quantum defects yielding smaller pair-defects for the intraspecies $d$-states than that of Cs. This also shows the advantage of using dual-species interactions, with the Rb-Cs coupling offering a strong blockaded interaction at $R=6~\mu$m where the intraspecies coupling is negligible.

For intermediate angles, Fig.~\ref{fig:potential_d_d}(e) and (f) show the pair-state eigenenergies as a function of $\theta$ calculated for a separation of $R=6~\mu$m. Unlike the characteristic $C_3\propto (1-3\cos^2(\theta))$ dependence obtained for single-species couplings between atoms in identical states \cite{ravets15}, the angular profiles show that operating more than around 5 degrees away from the optimal values of $\theta=0^\circ$ and $\pi/2$ the pair potentials show a complex energy landscape with a large number of additional eigenstates appearing that have weak or vanishing energy shifts, preventing realisation of strong blockaded interactions at these intermediate angles.

For the other interaction channels, a similar behavior is observed with stable resonances observed for $(m_j^\mathrm{Rb},m_j^\mathrm{Cs})=(+j^\mathrm{Rb},+j^\mathrm{Cs})$ at $\theta=0^\circ$ and $(+j^\mathrm{Rb},-j^\mathrm{Cs})$ at $90^\circ$. An example resonance for the Rb $59d_{5/2}$ - Cs $68d_{3/2}$ resonance at  $\theta=90^\circ$ and $(5/2,-3/2)$ is shown in Fig.~\ref{fig:compare}(a), which provides one of the best states for achieving low blockade and leakage errors as detailed later.

\subsection{Comparison to $s$-state \Forster resonances}
\begin{figure}[t]
\includegraphics{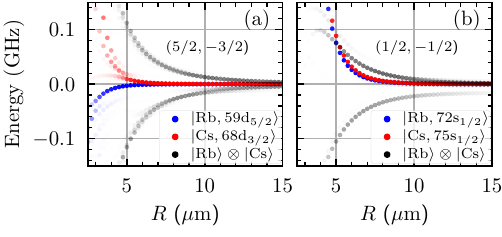}
\caption{Comparison of $d$ and $s$ state \Forster resonances with $\theta=90^\circ$ with magnetic sublevels denoted $(m_j^\mathrm{Rb},m_j^\mathrm{Cs})$. (a) Rb $59d_{5/2}$ - Cs $68d_{3/2}$ \Forster resonance for $(5/2,-3/2)$, and  (b) Rb $72s_{1/2}$ - Cs $75s_{1/2}$ \Forster resonance for $(1/2,-1/2)$. Interspecies pair-potentials are plotted black with opacity proportional to state overlap. Intraspecies interaction curves are shown for Rb (red) and Cs (blue).}
\label{fig:compare}
\end{figure}

To illustrate the advantage of the $d$-state resonances over $s$-states, Fig.~\ref{fig:compare}(b) shows an example \Forster resonance for the Rb $72s_{1/2}$ - Cs $70s_{1/2}$ state previously identified in Ref.~\cite{beterov15} for $\theta=90^\circ$ and $(m_j^\mathrm{Rb},m_j^\mathrm{Cs})=(1/2,-1/2)$. This reveals a strong interspecies resonance with comparable magnitude to that of Rb $59d_{5/2}$ - Cs $68d_{3/2}$ shown in Fig.~\ref{fig:compare}(a), but with the downside being that the intraspecies Rb-Rb and Cs-Cs interactions have a similar magnitude, negating the advantage of being able to exploit strong Rb-Cs interaction whilst suppressing the intraspecies couplings. 

A further benefit is that when performing two-photon excitation to the Rydberg states, the dipole-matrix elements to the intermediate state scale as $n^{-3/2}$. As well as obtaining comparable interspecies interactions at lower principal quantum numbers, the $d$-states have a pre-factor in this scaling approximately twice as strong as the $s$-states yielding enhanced Rabi frequencies at the same laser power.

\begin{table*}[t!]
    \centering
    \small
    \makebox[\linewidth]{
    \begin{tabular}{c|c|c|c|c|c||c|c||c|c}
        \hline\hline
         \multicolumn{2}{c|}{Rb - Cs \Forster Resonances}& $\Delta/2\pi$&$C_{3,k}$ & \multicolumn{2}{c||}{Rb-Cs at $R=6~\mu$m} & \multicolumn{2}{c||}{Rb-Rb at $R=8.5~\mu$m}  & \multicolumn{2}{c}{Cs-Cs at $R=8.5~\mu$m}\\
        Target Pair & Resonant Pair & (MHz) &(GHz$\,\mu$m$^3$) & $P_{1r}\left(t_\pi\right)$ & $\tilde{C}_3$ (GHz$\,\mu$m$^3$) & $P_{rr}\left(t_\pi\right)$ & $\tilde{C}_6$ (GHz$\,\mu$m$^6$) & $P_{rr}\left(t_\pi\right)$ & $\tilde{C}_6$ (GHz$\,\mu$m$^6$) \\         
        \hline \hline
	*$55d_{5/2}$ - $63d_{5/2}$ & $56p_{3/2}$ - $61f_{5/2}$ & 12.66 & -2.24 & 0.9969 & -11.30 & 0.9989 & 21.5 & 0.9840 & -98.8 \\
	$62d_{5/2}$ - $71d_{5/2}$ & $63p_{3/2}$ - $69f_{5/2}$ & -9.53 & -3.66 & 0.9997 & 15.79 & 0.9855 & 64.1 & 0.7747 & -416.8 \\
	$69d_{5/2}$ - $79d_{5/2}$ & $70p_{3/2}$ - $77f_{7/2}$ & -19.14 & -21.97 & 0.9998 & 26.34 & 0.7855 & 192.8 & 0.0013 & -1448.9 \\
	$70d_{5/2}$ - $80d_{5/2}$ & $71p_{3/2}$ - $78f_{7/2}$ & 13.69 & -23.21 & 0.9990 & -29.03 & 0.7442 & 223.9 & 0.0238 & -1812.1 \\
	$76d_{5/2}$ - $87d_{5/2}$ & $77p_{3/2}$ - $85f_{5/2}$ & -19.30 & -8.41 & 0.9997 & 40.58 & 0.1156 & 450.0 & 0.0388 & -4482.5 \\
	\hline 
	*$52d_{5/2}$ - $60d_{3/2}$ & $53p_{3/2}$ - $58f_{5/2}$ & 15.19 & -6.70 & 0.9964 & -8.37 & 0.9997 & 12.2 & 0.9958 & -45.0 \\
	*$59d_{5/2}$ - $68d_{3/2}$ & $60p_{3/2}$ - $66f_{5/2}$ & 13.40 & -11.23 & 0.9992 & -14.36 & 0.9951 & 41.6 & 0.9410 & -206.1 \\
	$66d_{5/2}$ - $76d_{3/2}$ & $67p_{3/2}$ - $74f_{5/2}$ & 11.12 & -17.74 & 0.9997 & -22.84 & 0.9176 & 114.4 & 0.4467 & -692.9 \\
	$79d_{5/2}$ - $91d_{3/2}$ & $80p_{3/2}$ - $89f_{5/2}$ & -12.53 & -36.95 & 0.9999 & 49.07 & 0.0070 & 619.2 & 0.0052 & -3596.5 \\
	$80d_{5/2}$ - $92d_{3/2}$ & $81p_{3/2}$ - $90f_{5/2}$ & 7.42 & -38.76 & 0.9999 & -51.12 & 0.0233 & 705.2 & 0.0144 & -5756.9 \\
	\hline
	*$53d_{3/2}$ - $60d_{5/2}$ & $54p_{1/2}$ - $58f_{5/2}$ & -4.41 & -1.97 & 0.9968 & 7.12 & 0.9989 & 19.7 & 0.9938 & -57.00 \\ 
	$76d_{3/2}$ - $87d_{5/2}$ & $77p_{3/2}$ - $85f_{5/2}$ & 6.49 & 2.81 & 0.9986 & -52.91 & 0.0129 & 723.9 & 0.0186 & -5038.3 \\
	\hline 
	$65d_{3/2}$ - $74d_{3/2}$ & $66p_{1/2}$ - $72f_{5/2}$ & -13.07 & -17.02 & 0.9974 & 16.67 & 0.7416 & 234.6 & 0.8297 & -375.1 \\
	$73d_{3/2}$ - $83d_{3/2}$ & $74p_{1/2}$ - $81f_{5/2}$ & -0.70 & -27.26 & 0.9995 & 28.54 & 0.1360 & 452.0 & 0.1082 & -2156.1 \\
	$80d_{3/2}$ - $91d_{3/2}$ & $81p_{1/2}$ - $89f_{5/2}$ & -13.66 & -39.62 & 0.9998 & 48.51 & 0.0440 & 1263.6 & 0.0234 & -3178.2 \\
	$81d_{3/2}$ - $92d_{3/2}$ & $82p_{1/2}$ - $90f_{5/2}$ & 4.39 & -41.55 & 0.9997 & -47.30 & 0.0590 & 1411.2 & 0.0153 & -3836.9 \\
	\hline 
	$72s_{1/2}$ - $70s_{1/2}$ & $71p_{3/2}$ - $70p_{1/2}$ & -7.90 & 8.51 & 0.9992 & 9.82 & 0.0040 & -1106.6 & 0.4005 & -605.3 \\
\hline\hline
    \end{tabular}}
    \caption{Summary of the strongest Rb-Cs $d$-state \Forster resonances offering $P_{1r}(t_\pi)\sim0.999$ for a pair of Rb-Cs atoms separated by $R=6~\mu$m calculated at $90^\circ$  for $(m_j^\mathrm{Rb},m_j^\mathrm{Cs})=(+j^\mathrm{Rb},-j^\mathrm{Cs})$assuming a single-photon Rabi frequency of $\Omega/2\pi=2$~MHz compared against the blockade errors expected for multi-qubit gates due to intraspecies coupling for Rb-Rb and Cs-Cs atoms at $8.5~\mu$m.  Along with the asymptotic pair energy defect $\Delta$ and channel coefficient $C_{3,k}$, we give the effective $\tilde{C}_3$ and $\tilde{C}_6$ coefficients obtained by fitting the  pair-potentials in the region $4<R<15~\mu$m. States labelled with an asterix $(*)$ are suitable for multi-qubit gate with weak intraspecies couplings. Also shown is the only $s$-state resonance acheiving $P_{1r}(t_\pi)>0.999$.}
    \label{tab:results}    
\end{table*}

\section{Blockade Leakage Errors}\label{sec:leakage}
In order to exploit Rydberg atom interactions for performing high-fidelity controlled gate operations, we use the blockade regime where the interaction strength $V$ greatly exceeds the excitation linewidth determined by the Rabi frequency $\Omega$ describing the rate of coupling from the computational ground-state to the target Rydberg state, preventing more than a single Rydberg excitation to be created.

For the \Forster resonances identified above, the exact pair potentials include coupling not just to a single level, but to many pair-eigenstates $\ket{\phi_i}$ each with energy $\epsilon_i$, and it is necessary to consider not only the blockade effect arising from the strongly shifted eigenstates but also blockade leakage due to coupling to weakly shifted eigenstates. To determine the blockade leakage error for a given target pair state $\ket{r_t^A,r_t^B}$, we calculate the probability of exciting this double Rydberg state following a resonant $\pi$-pulse applied to atom $A$ for atoms initialized at time $t=0$ in the state $\ket{\Psi(0)}=\ket{1^A,r_t^B}$, where $\ket{1^A}$ is the computational state of atom $A$, taking into account coupling to all possible eigenstates $\ket{\phi_i}$ found through diagonalizing the pair state interactions.

The Hamiltonian for the coupled system is given by
\begin{equation}\label{eq:analysis_H}
    \hat{H} = \hbar \sum_{i=1} \left( \left[ \frac{\eta_i \Omega}{2} \ket{\phi_i} \bra{0^A,r_t^B} + \text{h.c.} \right] - \epsilon_i \ket{\phi_i}\bra{\phi_i} \right),
\end{equation}
where $\Omega$ is the Rabi frequency for coupling $\ket{1^A}\rightarrow\ket{r_t^A}$ and $\eta_i=\bra{r_t^A,r_t^B}\phi_i\rangle$ denotes the relative overlap of the target pair state with the energy eigenstate, and the sum is taken for all eigenstates with $\vert \eta_i\vert>10^{-3}$. The resulting wavefunction at time $t$ is given by $\ket{\Psi(t)}=e^{-i\hat H t/\hbar}\ket{\Psi(0)}$, from which the probability of a double Rydberg excitation is given by 
\begin{equation}
    P_{rr}(t) = \sum_{i=1} \left| \braket{ \phi_i | \Psi(t) } \right|^2,
\end{equation}
and the probability of finding the atom in the initial ground state is $P_{1r}(t) = \left| \braket{ 1^A,r_t^B | \Psi(t) } \right|^2$.

For each of the possible \Forster-resonant pairs, we evaluate the blockade leakage error calculated for an Rb-Cs pair at a spacing of $R=6~\mu$m for $\theta=90^\circ$ and using $(+m_j^\mathrm{Rb},-m_j^\mathrm{Cs})$ following application of a resonant $\pi$-pulse for assuming a value of $\Omega/2\pi=2$~MHz. Results are presented in Tab.~\ref{tab:results} which is filtered to select states with $P_{1r}(t_\pi)>0.999$ consistent with strong blockade and leakage errors below $10^{-3}$. For the Rb $d_{3/2}$-Cs $d_{5/2}$ interactions no pair states reach this limit, with two example resonances included for completeness, whilst for the other interaction channels several candidate pair states are identified with leakage errors as low as $10^{-4}$ for Rb $79d_{5/2}$-Cs $91d_{3/2}$ and Rb $80d_{5/2}$-Cs $92d_{3/2}$.

Since one of the motivations for using dual species interactions is not only to obtain strong Rb-Cs couplings, but also to provide a route to realising multi-qubit gates suppressing target-target and control-control couplings, we additionally consider the impact of intraspecies interactions. We assume atoms are arranged on a square lattice as shown in Fig.~\ref{fig:schematic}, and calculate the probability of creating doubly excited Rb-Rb and Cs-Cs pair states at a minimum distance of $\sqrt{2}R=8.5~\mu$m. For these states we evaluate both the probability of creating a double excitation after a $\pi$-pulse, and of returning back to the initial state following a $2\pi$ pulse which in the ideal case of zero intraspecies interactions would return unity in both cases. We also tabulate the effective $\tilde{C}_3$ and $\tilde{C}_6$ coefficients obtained by fitting the dominant interaction curves in the region $4<R<15~\mu$m for the Rb-Cs, Rb-Rb and Cs-Cs interaction curves to enable extrapolation of the interaction strength within these distances typically used for tweezer experiments.

From Table~\ref{tab:results} it can be seen that whilst many suitable Rb-Cs \Forster resonances can be found, those offering the highest suppression of blockade leakage typically correspond to states with higher principal quantum number ($n\gtrsim65$) where the residual intraspecies interactions at 8.5~$\mu$m are sufficiently strong that there is a signficant blockade effect visible for the Rb-Rb and Cs-Cs curves, making these poorly suited to extension to multi-qubit gates. Whilst in practise these errors can be suppressed to some extent by using increased single-photon Rabi frequencies, this comes at the cost of increased leakage errors for the Rb-Cs interactions. This also highlights the limitations of using $n_\mathrm{Rb}d_{3/2}$ states above, as from the table it can be seen that for $n_\mathrm{Rb}\ge58$ these have much a much stronger blockade error for Rb-Rb than the comparable $n_\mathrm{Rb}d_{5/2}$ states. For completeness we note only a single $s$-state resonance achieves $P_{1r}(t_\pi)>0.999$ for which the resulting Rb-Rb and Cs-Cs interactions are too strongly blockaded for use for multi-qubit gates in agreement with the pair-potential shown above in Fig.~\ref{fig:compare}(b).

Imposing the requirement to not only obtain low leakage errors for Rb-Cs but also reduced errors for Rb-Rb and Cs-Cs operations, we identify a number of possible states denoted by an asterisk in the table that balance the different error channels. From these candidates $55d_{5/2}63d_{5/2}$, $52d_{5/2}60d_{3/2}$, and $53d_{3/2}60d_{5/2}$ offer comparable performance of $P>0.99$ for all metrics, and $59d_{5/2}68d_{3/2}$ achieves the lowest intraspecies error with $P_{rr}(t_\pi)=0.995$ for Rb-Rb whilst achieving $P_{1r}(t_\pi)>0.999$ for Rb-Cs. In the next section we use this state to estimate realistic multi-qubit gate performance.

\section{Dual-species multi-qubit gate operations}
\label{sec:gate}
In the analysis above the only error considered in performing gate operations has been that of blockade errors due to errors associated with leakage of population into doubly-occupied states for the Rb-Cs interactions, and the suppression of doubly occupied states for the Rb-Rb and Cs-Cs case due to residual intraspecies couplings. To accurately estimate gate fidelity, it is necessary to also include errors associated with spontaneous emission from the Rydberg state as well as scattering from the intermediate excited state during the two-photon excitation of atoms to the Rydberg state.

To efficiently simulate these errors, we adopt the effective model introduced in Ref.~\cite{pelegri22} to model each atom using three computational levels with an effective single-atom Hamiltonian describing atom-light interactions of the form
\begin{eqnarray}\label{eq:Heff}
    \hat{H}_\mathrm{eff} = \frac{\hbar}{2}\Bigl[&-iP_e^1\Gamma_e\ket{1}\bra{1}+
    \Omega_R(\ket{r}\bra{1}+\ket{1}\bra{r})\nonumber\\
    &-(\delta+i\Gamma_r+P_e^r\Gamma_e)\ket{r}\bra{r}\Bigr],
\end{eqnarray}
where $\Omega_R$ is the effective two-photon Rabi frequency, $\delta$ is the two-photon detuning, $\Gamma_{e,r}$ are the excited and Rydberg state linewidths and $P_e^1$ and $P_e^r$ are excited state probabilities determined from performing adiabatic elimination of the intermediate excited states. For gates performed using the Rb $59d_{5/2}$-Cs $68d_{3/2}$ \Forster{} resonance, we first determine suitable parameters for the two-photon Rabi frequency to ensure atoms obtain a $\pi$-phase shift when excited through $2\pi$ by tuning the ratio between the two-photon Rabi frequencies \cite{maller15}. For both species, we assume a single-photon detuning of 10~GHz from the excited state transition and determine parameters giving $\Omega_R/2\pi=5$~MHz. For Rb, excitations are modelled via the $6p_{3/2}$ intermediate state with $\Gamma_e/2\pi=1.4$~MHz and for Cs via $7p_{1/2}$ with $\Gamma_e/2\pi=1.0$~MHz.  Full details of the excitation parameters are given in Appendix~\ref{sec:details}.

Using the effective Hamiltonian, we model the canonical phase-gate protocol based on a $\pi$-pulse on the control qubit, followed by a $2\pi$-pulse on the target and a third $\pi$-pulse on the control qubit for an initial symmetric superposition of all possible input states. We note that whilst this is not an optimal pulse sequence, it provides a simple approach to extending to $C^kZ$ gates yielding the correct output states up to single qubit rotations \cite{levine19}. For these diagonal controlled phase-gates, this also enables efficient estimation of the resulting gate fidelity by calculating $\mathcal{F}=\vert \bra{\Psi_I}\Psi\rangle\vert^2$ where $\ket{\Psi}$ is the final wavefunction and $\ket{\Psi_I}$ is the output of an ideal gate.

We first consider a two-qubit $CZ$ gate for Rb-Cs atoms separated by $R=6~\mu$m, which experience an interaction shift of $V_\mathrm{Rb-Cs}/2\pi=65.6$~MHz. Using Rb as the control qubit give $\mathcal{F}_{CZ}=0.9953$, whilst for Cs as the control qubit $\mathcal{F}_{CZ}=0.9955$, competitive against recent experimental demonstrations \cite{evered23,ma23}. Scaling to multi-qubit gates, we recall that this pair-state was chosen due to the suppressed Rb-Rb interactions and proceed to use a single Cs target atom and increase the number of neighboring Rb control atoms.

For the $CCZ$ gate, placing atoms on a line with the Rb control atoms either side of a central Cs target, the Rb-Rb interaction at 16~$\mu$m corresponds to $V_\mathrm{Rb-Rb}/2\pi=0.11$~MHz, resulting in $\mathcal{F}_{CCZ}=0.994$, whilst moving the second Rb control atom to the site below the Cs (at a reduced Rb-Rb distance of $8.5~\mu$m) increases the control-control coupling to $V_\mathrm{Rb-Rb}/2\pi=0.14$~MHz and results in $\mathcal{F}_{CCZ}=0.983$. Extending to a $C^3Z$ gate with three Rb atoms in an equilateral triangle configuration with the Cs atom at the centre gives $\mathcal{F}_{C^3Z}=0.988$, and for the square geometry shown in Fig.~\ref{fig:schematic}(c) we find a $C^4Z$ gate gives $\mathcal{F}_{C^4Z}=0.913$. 

Whilst even with the engineered interspecies interactions we still observe a reduction in fidelity due to residual intraspecies coupling, decomposition of the $CCZ$ gate requires 6 $CZ$ operations \cite{shende08}, which would require $\mathcal{F}_{CZ}>0.999$ and $0.997$ respectively for the line and square configurations, demonstrating the benefit of using dual-species couplings for native multi-qubit gate implementations. Additionally, further optimizaton of gate pulses taking account of additional technical noise sources could yield higher gate performance, but lies beyond the scope of this current work.


\section{Conclusion}\label{sec:conclusion}
In this paper we have presented a detailed analysis of the $d$-state \Forster resonances for engineering strong Rb-Cs interactions to enable high-fidelity gate operations between heterogeneous atom pairs. We have identified dominant interaction channels yielding strong $C_{3,k}$ coefficients and small pair defects, and demonstrated the importance of performing diagonalization of the pair-state potentials to identify suitable resonances for achieving strong short-range interactions. 

From studying the angular dependence of the pair potentials we have shown that for the $d$-states the \Forster resonances are only robust at $\theta=0^\circ$ or $90^\circ$ dependent upon choice of magnetic sublevel, with $(+m_j^\mathrm{Rb},-m_j^\mathrm{Cs})$ providing the best performance at $\theta=90^\circ$, but with additional benefits compared to the $s$-state resonances through much greater suppression of the intra-species interactions and providing stronger resonant interactions at lower principal quantum number.

By calculating blockade leakage and excitation errors using the full pair-state eigenvalues, we have identified states offering the best performance for realising robust blockade for controlled gate operations, and demonstrated the ability to use these strong interspecies couplings to realize high-fidelity $C^kZ$ gates due to suppression of the residual intraspecies couplings between control qubits compared to single-species gate implementations, providing a route to native multi-qubit gates with performance exceeding that expected for the equivalent two-qubit gate decompositions using current best-known protocols \cite{evered23,pelegri22}.

These results highlight the potential advantages of developing scalable neutral atom processors based on dual atomic species, and will guide future demonstrations of gate operations on these emerging Rb-Cs platforms \cite{singh22,guttridge23} by providing improved choice over the best \Forster-resonant pair states to target.

\begin{acknowledgments}
This work was funded by the Royal Academy of Engineering Senior Research Fellowship with M Squared Lasers and the Engineering and Physical Sciences Research Council (Grant No. EP/X025055/1). The data presented in the paper are available here \cite{ireland23data}.
\end{acknowledgments}


%

\newpage{}
\appendix

\section{Tables of Rb-Cs $d$-state \Forster{} Resonances}
Below in Tables~\ref{tab:DD} to \ref{tab:dD} we present $n_\mathrm{Rb}d_j$-$n_\mathrm{Cs}d_j$ \Forster resonant pair states with $\vert \Delta\vert/2\pi\lesssim20$~MHz and $\vert C_{3,k}\vert>1~$GHz$\,\mu$m$^3$ for the different combination of $j=3/2,5/2$. For each state we additionally show $P_{1r}(t_\pi)$ for Rb-Cs at $R=6~\mu$m to quantify blockade leakage errors, and  $P_{rr}(t_\pi)$ for Rb-Rb and Cs-Cs interactions at $R=8.5~\mu$m representative of expected errors for dual-species multi-qubit gates on a square grid. Finally, we include the dispersion coefficients $\tilde{C}_3$ and $\tilde{C}_6$ for inter- and intra-species couplings respectively obtained by fitting the pair-eigenstate with the largest target state probability in the region $5\le R \le15~\mu$m. For comparison Tab.~\ref{tab:ss} includes the parameters obtained for $s$-state \Forster-resonant pairs previously identified in Ref.~\cite{beterov15}, which typically have higher leakage errors than the $d$-states and much stronger blockade error for intraspecies couplings with similar magnitude and range as the Rb-Cs coupling.

\begin{table*}[b!]
    \centering
    \small
    \makebox[\linewidth]{
    \begin{tabular}{c|c|c|c|c|c||c|c||c|c}
        \hline\hline
         \multicolumn{2}{c|}{Rb - Cs \Forster Resonances}& $\Delta/2\pi$&$C_{3,k}$ & \multicolumn{2}{c||}{Rb-Cs at $R=6~\mu$m} & \multicolumn{2}{c||}{Rb-Rb at $R=8.5~\mu$m}  & \multicolumn{2}{c}{Cs-Cs at $R=8.5~\mu$m}\\
        Target Pair & Resonant Pair & (MHz) &(GHz$\,\mu$m$^3$) & $P_{1r}\left(t_\pi\right)$ & $\tilde{C}_3$ (GHz$\,\mu$m$^3$) & $P_{rr}\left(t_\pi\right)$ & $\tilde{C}_6$ (GHz$\,\mu$m$^6$) & $P_{rr}\left(t_\pi\right)$ & $\tilde{C}_6$ (GHz$\,\mu$m$^6$) \\         
        \hline \hline
	$55d_{5/2}$ - $53d_{5/2}$ & $57p_{3/2}$ - $50f_{7/2}$ & 18.71 & -2.73 & 0.9860 & -3.13 & 0.9989 & 21.4 & 0.9999 & -7.4 \\
	$55d_{5/2}$ - $63d_{5/2}$ & $56p_{3/2}$ - $61f_{5/2}$ & 12.66 & -2.24 & 0.9969 & -11.30 & 0.9989 & 21.5 & 0.9840 & -98.8 \\
	$55d_{5/2}$ - $61d_{5/2}$ & $53f_{7/2}$ - $63p_{3/2}$ & -12.19 & -1.78 & 0.9747 & 2.08 & 0.9989 & 21.4 & 0.9923 & -62.1 \\
	$62d_{5/2}$ - $71d_{5/2}$ & $63p_{3/2}$ - $69f_{5/2}$ & -9.53 & -3.66 & 0.9997 & 15.79 & 0.9855 & 64.1 & 0.7747 & -416.8 \\
	$69d_{5/2}$ - $79d_{5/2}$ & $70p_{3/2}$ - $77f_{7/2}$ & -19.14 & -21.97 & 0.9998 & 26.34 & 0.7855 & 192.8 & 0.0013 & -1448.9 \\
	$69d_{5/2}$ - $66d_{5/2}$ & $71p_{3/2}$ - $63f_{5/2}$ & -4.29 & -1.79 & 0.9933 & 7.49 & 0.7855 & 192.8 & 0.9538 & -164.5 \\
	$77d_{5/2}$ - $88d_{5/2}$ & $78p_{3/2}$ - $86f_{7/2}$ & 1.91 & -34.24 & 0.9998 & -47.42 & 0.2966 & 393.5 & 0.0282 & -4303.1 \\
	$74d_{5/2}$ - $82d_{5/2}$ & $72f_{7/2}$ - $84p_{3/2}$ & 5.63 & -6.15 & 0.9967 & -16.85 & 0.3269 & 367.5 & 0.1115 & -2258.4 \\
	$70d_{5/2}$ - $80d_{5/2}$ & $71p_{3/2}$ - $78f_{7/2}$ & 13.69 & -23.21 & 0.9990 & -29.03 & 0.7442 & 223.9 & 0.0238 & -1812.1 \\
	$76d_{5/2}$ - $87d_{5/2}$ & $77p_{3/2}$ - $85f_{5/2}$ & -19.30 & -8.41 & 0.9997 & 40.58 & 0.1156 & 450.0 & 0.0388 & -4482.5 \\
	$84d_{5/2}$ - $80d_{5/2}$ & $86p_{3/2}$ - $77f_{7/2}$ & 11.74 & -15.58 & 0.9996 & -24.33 & 0.0514 & 1318.7 & 0.0148 & -1692.0 \\
	$80d_{5/2}$ - $70d_{5/2}$ & $78f_{7/2}$ - $68f_{5/2}$ & -18.70 & 3.05 & 0.9932 & 2.38 & 0.0229 & 698.0 & 0.8301 & -356.6 \\
	$83d_{5/2}$ - $79d_{5/2}$ & $85p_{3/2}$ - $76f_{7/2}$ & -10.78 & -14.80 & 0.9975 & 12.46 & 0.0957 & 1108.5 & 0.0013 & -1448.9 \\
\hline\hline
    \end{tabular}}
    \caption{Rb-Cs \Forster resonances from Rb $d_{5/2}$ - Cs $d_{5/2}$ states with $\vert \Delta\vert /2\pi<20~$MHz and $\vert C_{3,k}\vert>1$~GHz$\,\mu$m$^3$. $P_{1r}(t_\pi)$ and $P_{rr}(t_\pi)$ along with fitted $\tilde{C}_3$ and $\tilde{C}_6$ coefficients are calculated for $\theta=90^\circ$ and $(m_j^\mathrm{Rb},m_j^\mathrm{Cs})=(5/2,-5/2)$ assuming $\Omega/2\pi=2$~MHz.}
    \label{tab:DD}
\end{table*}
\begin{table*}[t!]
    \centering
    \small
    \makebox[\linewidth]{
    \begin{tabular}{c|c|c|c|c|c||c|c||c|c}
        \hline\hline
         \multicolumn{2}{c|}{Rb - Cs \Forster Resonances}& $\Delta/2\pi$&$C_{3,k}$ & \multicolumn{2}{c||}{Rb-Cs at $R=6~\mu$m} & \multicolumn{2}{c||}{Rb-Rb at $R=8.5~\mu$m}  & \multicolumn{2}{c}{Cs-Cs at $R=8.5~\mu$m}\\
        Target Pair & Resonant Pair & (MHz) &(GHz$\,\mu$m$^3$) & $P_{1r}\left(t_\pi\right)$ & $\tilde{C}_3$ (GHz$\,\mu$m$^3$) & $P_{rr}\left(t_\pi\right)$ & $\tilde{C}_6$ (GHz$\,\mu$m$^6$) & $P_{rr}\left(t_\pi\right)$ & $\tilde{C}_6$ (GHz$\,\mu$m$^6$) \\         
        \hline \hline
	$51d_{3/2}$ - $59d_{3/2}$ & $52p_{3/2}$ - $57f_{5/2}$ & -18.12 & 2.08 & 0.9912 & 0.83 & 0.9995 & 13.8 & 0.9995 & -15.8 \\
	$50d_{3/2}$ - $57d_{3/2}$ & $51p_{1/2}$ - $55f_{5/2}$ & 16.13 & -5.82 & 0.9958 & -6.33 & 0.9997 & 11.4 & 0.9998 & -10.9 \\
	$51d_{3/2}$ - $62d_{3/2}$ & $50f_{5/2}$ - $59f_{5/2}$ & -15.98 & 4.52 & 0.9814 & 3.15 & 0.9995 & 13.8 & 0.9995 & -13.2 \\
	$65d_{3/2}$ - $75d_{3/2}$ & $66p_{3/2}$ - $73f_{5/2}$ & 8.32 & 5.58 & 0.9955 & -36.55 & 0.7416 & 234.6 & 0.7704 & -440.2 \\
	$69d_{3/2}$ - $66d_{3/2}$ & $71p_{1/2}$ - $63f_{5/2}$ & -6.16 & -7.35 & 0.9979 & 6.81 & 0.4719 & 224.4 & 0.9706 & -141.0 \\
	$64d_{3/2}$ - $61d_{3/2}$ & $66p_{3/2}$ - $58f_{5/2}$ & 3.96 & 1.64 & 0.8913 & -0.35 & 0.7767 & 211.9 & 0.9939 & -63.6 \\
	$65d_{3/2}$ - $74d_{3/2}$ & $66p_{1/2}$ - $72f_{5/2}$ & -13.07 & -17.02 & 0.9974 & 16.67 & 0.7416 & 234.6 & 0.8297 & -375.1 \\
	$76d_{3/2}$ - $67d_{3/2}$ & $74f_{5/2}$ - $65f_{5/2}$ & 2.04 & 8.99 & 0.9868 & -12.99 & 0.0129 & 723.9 & 0.9587 & -163.3 \\
	$77d_{3/2}$ - $73d_{3/2}$ & $79p_{3/2}$ - $70f_{5/2}$ & -6.12 & 3.50 & 0.9789 & 3.82 & 0.0692 & 820.6 & 0.8729 & -308.5 \\
	$71d_{3/2}$ - $79d_{3/2}$ & $69f_{5/2}$ - $81p_{3/2}$ & 9.09 & 1.63 & 0.9944 & -11.82 & 0.2858 & 329.2 & 0.3238 & -545.4 \\
	$70d_{3/2}$ - $77d_{3/2}$ & $68f_{5/2}$ - $79p_{1/2}$ & 7.97 & -5.30 & 0.9961 & -10.84 & 0.3789 & 287.4 & 0.6138 & -629.1 \\
	$72d_{3/2}$ - $83d_{3/2}$ & $73p_{3/2}$ - $81f_{5/2}$ & 10.33 & 8.46 & 0.9985 & -42.76 & 0.2049 & 390.5 & 0.1082 & -2156.1 \\
	$79d_{3/2}$ - $91d_{3/2}$ & $80p_{3/2}$ - $89f_{5/2}$ & 10.38 & 12.33 & 0.9998 & -51.08 & 0.1280 & 1112.0 & 0.0234 & -3178.2 \\
	$78d_{3/2}$ - $90d_{3/2}$ & $79p_{3/2}$ - $88f_{5/2}$ & -10.66 & 11.75 & 0.9998 & 21.39 & 0.1607 & 957.1 & 0.0262 & -2568.1 \\
	$73d_{3/2}$ - $83d_{3/2}$ & $74p_{1/2}$ - $81f_{5/2}$ & -0.70 & -27.26 & 0.9995 & 28.54 & 0.1360 & 452.0 & 0.1082 & -2156.1 \\
	$81d_{3/2}$ - $92d_{3/2}$ & $82p_{1/2}$ - $90f_{5/2}$ & 4.39 & -41.55 & 0.9997 & -47.30 & 0.0590 & 1411.2 & 0.0153 & -3836.9 \\
	$84d_{3/2}$ - $74d_{3/2}$ & $82f_{5/2}$ - $72f_{5/2}$ & -8.89 & 13.56 & 0.9906 & 3.10 & 0.0368 & 1937.6 & 0.6502 & -484.5 \\
	$84d_{3/2}$ - $80d_{3/2}$ & $86p_{1/2}$ - $77f_{5/2}$ & 11.47 & -16.47 & 0.9953 & -24.02 & 0.0368 & 1937.6 & 0.7637 & -388.1 \\
	$80d_{3/2}$ - $89d_{3/2}$ & $78f_{5/2}$ - $91p_{3/2}$ & 4.98 & 2.68 & 0.9966 & -21.96 & 0.0440 & 1263.6 & 0.0135 & -2640.7 \\
	$80d_{3/2}$ - $88d_{3/2}$ & $78f_{5/2}$ - $90p_{1/2}$ & -3.06 & -9.23 & 0.9357 & 9.12 & 0.0440 & 1263.6 & 0.0007 & -2226.4 \\
	$80d_{3/2}$ - $91d_{3/2}$ & $81p_{1/2}$ - $89f_{5/2}$ & -13.66 & -39.62 & 0.9998 & 48.51 & 0.0440 & 1263.6 & 0.0234 & -3178.2 \\
	$83d_{3/2}$ - $79d_{3/2}$ & $85p_{1/2}$ - $76f_{5/2}$ & -10.65 & -15.65 & 0.9983 & 10.75 & 0.0603 & 1610.2 & 0.0904 & -1035.9 \\
	\hline \\
\hline\hline
    \end{tabular}}
\caption{Rb-Cs \Forster resonances from Rb $d_{3/2}$ - Cs $d_{3/2}$ states with $\vert \Delta\vert /2\pi<20~$MHz and $\vert C_{3,k}\vert>1$~GHz$\,\mu$m$^3$. $P_{1r}(t_\pi)$ and $P_{rr}(t_\pi)$ along with fitted $\tilde{C}_3$ and $\tilde{C}_6$ coefficients are calculated for $\theta=90^\circ$ and $(m_j^\mathrm{Rb},m_j^\mathrm{Cs})=(3/2,-3/2)$ assuming $\Omega/2\pi=2$~MHz.}
    \label{tab:dd}
\end{table*}

\begin{table*}[t!]
    \centering
    \small
    \makebox[\linewidth]{
    \begin{tabular}{c|c|c|c|c|c||c|c||c|c}
        \hline\hline
         \multicolumn{2}{c|}{Rb - Cs \Forster Resonances}& $\Delta/2\pi$&$C_{3,k}$ & \multicolumn{2}{c||}{Rb-Cs at $R=6~\mu$m} & \multicolumn{2}{c||}{Rb-Rb at $R=8.5~\mu$m}  & \multicolumn{2}{c}{Cs-Cs at $R=8.5~\mu$m}\\
        Target Pair & Resonant Pair & (MHz) &(GHz$\,\mu$m$^3$) & $P_{1r}\left(t_\pi\right)$ & $\tilde{C}_3$ (GHz$\,\mu$m$^3$) & $P_{rr}\left(t_\pi\right)$ & $\tilde{C}_6$ (GHz$\,\mu$m$^6$) & $P_{rr}\left(t_\pi\right)$ & $\tilde{C}_6$ (GHz$\,\mu$m$^6$) \\         
        \hline \hline
	$59d_{5/2}$ - $68d_{3/2}$ & $60p_{3/2}$ - $66f_{5/2}$ & 13.40 & -11.23 & 0.9992 & -14.36 & 0.9951 & 41.6 & 0.9410 & -206.1 \\
	$50d_{5/2}$ - $37d_{3/2}$ & $49f_{7/2}$ - $38p_{1/2}$ & -5.91 & -2.93 & 0.9864 & 3.87 & 0.9999 & 8.2 & 1.0000 & -0.2 \\
	$52d_{5/2}$ - $60d_{3/2}$ & $53p_{3/2}$ - $58f_{5/2}$ & 15.19 & -6.70 & 0.9964 & -8.37 & 0.9997 & 12.2 & 0.9958 & -45.0 \\
	$66d_{5/2}$ - $76d_{3/2}$ & $67p_{3/2}$ - $74f_{5/2}$ & 11.12 & -17.74 & 0.9997 & -22.84 & 0.9176 & 114.4 & 0.4467 & -692.9 \\
	$68d_{5/2}$ - $60d_{3/2}$ & $66f_{5/2}$ - $58f_{5/2}$ & -8.50 & 1.52 & 0.9272 & 3.71 & 0.8373 & 170.4 & 0.9993 & -18.9 \\
	$60d_{5/2}$ - $66d_{3/2}$ & $58f_{7/2}$ - $68p_{1/2}$ & -14.00 & -2.87 & 0.9906 & 2.65 & 0.9929 & 48.3 & 0.9698 & -144.1 \\
	$67d_{5/2}$ - $49d_{3/2}$ & $66f_{7/2}$ - $50p_{1/2}$ & -5.80 & -9.64 & 0.9932 & 11.22 & 0.8889 & 142.5 & 1.0000 & -2.7 \\
	$78d_{5/2}$ - $74d_{3/2}$ & $80p_{3/2}$ - $71f_{5/2}$ & 2.73 & -11.14 & 0.9977 & -16.52 & 0.0056 & 557.1 & 0.8297 & -375.1 \\
	$79d_{5/2}$ - $91d_{3/2}$ & $80p_{3/2}$ - $89f_{5/2}$ & -12.53 & -36.95 & 0.9999 & 49.07 & 0.0070 & 619.2 & 0.0052 & -3596.5 \\
	$73d_{5/2}$ - $84d_{3/2}$ & $74p_{3/2}$ - $82f_{5/2}$ & 9.09 & -26.72 & 0.9949 & -36.07 & 0.4852 & 314.6 & 0.1085 & -2230.7 \\
	$77d_{5/2}$ - $56d_{3/2}$ & $76f_{7/2}$ - $57p_{1/2}$ & -20.99 & -16.90 & 0.9912 & 17.86 & 0.2966 & 393.5 & 0.9996 & -14.4 \\
	$81d_{5/2}$ - $90d_{3/2}$ & $79f_{7/2}$ - $92p_{3/2}$ & 16.57 & 2.90 & 0.9953 & -19.25 & 0.0649 & 765.4 & 0.0024 & -3282.6 \\
	$83d_{5/2}$ - $73d_{3/2}$ & $81f_{5/2}$ - $71f_{5/2}$ & 15.70 & 3.43 & 0.9939 & -12.68 & 0.0957 & 1108.5 & 0.8729 & -308.5 \\
	$81d_{5/2}$ - $89d_{3/2}$ & $79f_{5/2}$ - $91p_{1/2}$ & 5.56 & -2.58 & 0.9802 & -23.45 & 0.0858 & 839.1 & 0.0261 & -2866.1 \\
	$84d_{5/2}$ - $61d_{3/2}$ & $86f_{7/2}$ - $61p_{3/2}$ & -6.70 & -0.00 & 0.9991 & 26.62 & 0.0774 & 1227.1 & 0.9939 & -63.6 \\
	$80d_{5/2}$ - $92d_{3/2}$ & $81p_{3/2}$ - $90f_{5/2}$ & 7.42 & -38.76 & 0.9999 & -51.12 & 0.0233 & 705.2 & 0.0144 & -5756.9 \\
\hline\hline
    \end{tabular}}
    \caption{Rb-Cs \Forster resonances from Rb $d_{5/2}$ - Cs $d_{3/2}$ states with $\vert \Delta\vert /2\pi\lesssim20~$MHz and $\vert C_{3,k}\vert>1$~GHz$\,\mu$m$^3$. $P_{1r}(t_\pi)$ and $P_{rr}(t_\pi)$ along with fitted $\tilde{C}_3$ and $\tilde{C}_6$ coefficients are calculated for $\theta=90^\circ$ and $(m_j^\mathrm{Rb},m_j^\mathrm{Cs})=(5/2,-3/2)$ assuming $\Omega/2\pi=2$~MHz.}
    \label{tab:Dd}
\end{table*}

\begin{table*}[t!]
    \centering
    \small
    \makebox[\linewidth]{
    \begin{tabular}{c|c|c|c|c|c||c|c||c|c}
        \hline\hline
         \multicolumn{2}{c|}{Rb - Cs \Forster Resonances}& $\Delta/2\pi$&$C_{3,k}$ & \multicolumn{2}{c||}{Rb-Cs at $R=6~\mu$m} & \multicolumn{2}{c||}{Rb-Rb at $R=8.5~\mu$m}  & \multicolumn{2}{c}{Cs-Cs at $R=8.5~\mu$m}\\
        Target Pair & Resonant Pair & (MHz) &(GHz$\,\mu$m$^3$) & $P_{1r}\left(t_\pi\right)$ & $\tilde{C}_3$ (GHz$\,\mu$m$^3$) & $P_{rr}\left(t_\pi\right)$ & $\tilde{C}_6$ (GHz$\,\mu$m$^6$) & $P_{rr}\left(t_\pi\right)$ & $\tilde{C}_6$ (GHz$\,\mu$m$^6$) \\         
        \hline \hline
	$59d_{3/2}$ - $57d_{5/2}$ & $61p_{1/2}$ - $54f_{5/2}$ & 2.93 & -1.03 & 0.9823 & -5.56 & 0.6608 & 187.8 & 0.9981 & -32.2 \\
	$53d_{3/2}$ - $60d_{5/2}$ & $54p_{1/2}$ - $58f_{5/2}$ & -4.41 & -1.97 & 0.9968 & 7.12 & 0.9989 & 19.7 & 0.9938 & -57.0 \\
	$64d_{3/2}$ - $71d_{5/2}$ & $62f_{5/2}$ - $73p_{3/2}$ & 4.08 & -3.26 & 0.9909 & -7.31 & 0.7767 & 211.9 & 0.7383 & -435.9 \\
	$78d_{3/2}$ - $88d_{5/2}$ & $79p_{1/2}$ - $86f_{5/2}$ & 6.51 & -9.49 & 0.9915 & -39.56 & 0.1607 & 957.1 & 0.0085 & -5492.7 \\
	$76d_{3/2}$ - $87d_{5/2}$ & $77p_{3/2}$ - $85f_{5/2}$ & 6.49 & 2.81 & 0.9986 & -52.91 & 0.0129 & 723.9 & 0.0186 & -5038.3 \\
	$80d_{3/2}$ - $70d_{5/2}$ & $78f_{5/2}$ - $68f_{5/2}$ & 3.69 & 2.95 & 0.9920 & -11.62 & 0.0440 & 1263.6 & 0.8048 & -377.9 \\
	$83d_{3/2}$ - $79d_{5/2}$ & $85p_{3/2}$ - $76f_{7/2}$ & 8.93 & 4.92 & 0.9918 & -32.33 & 0.0522 & 1813.4 & 0.0000 & -1466.9 \\
	$90d_{3/2}$ - $86d_{5/2}$ & $92p_{1/2}$ - $83f_{5/2}$ & 2.55 & -5.81 & 0.9961 & -40.04 & 0.0066 & 3762.9 & 0.0201 & -4212.1 \\
    \hline\hline
    \end{tabular}}
    \caption{Rb-Cs \Forster resonances from Rb $d_{3/2}$ - Cs $d_{5/2}$ states with $\vert \Delta\vert /2\pi\lesssim20~$MHz and $\vert C_{3,k}\vert>1$~GHz$\,\mu$m$^3$. $P_{1r}(t_\pi)$ and $P_{rr}(t_\pi)$ along with fitted $\tilde{C}_3$ and $\tilde{C}_6$ coefficients are calculated for $\theta=90^\circ$ and $(m_j^\mathrm{Rb},m_j^\mathrm{Cs})=(3/2,-5/2)$ assuming $\Omega/2\pi=2$~MHz.}
    \label{tab:dD}
\end{table*}

\begin{table*}[t!]
    \centering
    \small
    \makebox[\linewidth]{
    \begin{tabular}{c|c|c|c|c|c||c|c||c|c}
        \hline\hline
         \multicolumn{2}{c|}{Rb - Cs \Forster Resonances}& $\Delta/2\pi$&$C_{3,k}$ & \multicolumn{2}{c||}{Rb-Cs at $R=6~\mu$m} & \multicolumn{2}{c||}{Rb-Rb at $R=8.5~\mu$m}  & \multicolumn{2}{c}{Cs-Cs at $R=8.5~\mu$m}\\
        Target Pair & Resonant Pair & (MHz) &(GHz$\,\mu$m$^3$) & $P_{1r}\left(t_\pi\right)$ & $\tilde{C}_3$ (GHz$\,\mu$m$^3$) & $P_{rr}\left(t_\pi\right)$ & $\tilde{C}_6$ (GHz$\,\mu$m$^6$) & $P_{rr}\left(t_\pi\right)$ & $\tilde{C}_6$ (GHz$\,\mu$m$^6$) \\         
        \hline \hline
	$48s_{1/2}$ - $51s_{1/2}$ & $48p_{3/2}$ - $50p_{1/2}$ & -5.53 & 1.69 & 0.6281 & 3.00 & 0.9998 & -9.5 & 0.9996 & -14.5 \\
	$59s_{1/2}$ - $57s_{1/2}$ & $58p_{1/2}$ - $57p_{1/2}$ & -16.45 & 3.54 & 0.3752 & 1.35 & 0.9767 & -114.2 & 0.9943 & -56.6 \\
	$69s_{1/2}$ - $68s_{1/2}$ & $68p_{1/2}$ - $68p_{3/2}$ & -9.89 & 6.92 & 0.9959 & 12.69 & 0.3230 & -701.2 & 0.6503 & -430.2 \\
	$68s_{1/2}$ - $67s_{1/2}$ & $67p_{1/2}$ - $67p_{3/2}$ & 2.63 & 6.50 & 0.9727 & -9.12 & 0.4660 & -594.3 & 0.7432 & -367.5 \\
	$61s_{1/2}$ - $65s_{1/2}$ & $61p_{1/2}$ - $64p_{1/2}$ & 2.77 & 4.80 & 0.9735 & -2.62 & 0.9487 & -168.9 & 0.8689 & -263.5 \\
	$72s_{1/2}$ - $75s_{1/2}$ & $72p_{1/2}$ - $74p_{3/2}$ & 2.70 & 9.65 & 0.9835 & -12.40 & 0.0040 & -1106.6 & 0.0359 & -1187.3 \\
	$77s_{1/2}$ - $81s_{1/2}$ & $77p_{3/2}$ - $80p_{1/2}$ & -2.10 & 12.28 & 0.9969 & 16.42 & 0.0100 & -2103.4 & 0.0344 & -2151.8 \\
	$72s_{1/2}$ - $70s_{1/2}$ & $71p_{3/2}$ - $70p_{1/2}$ & -7.90 & 8.51 & 0.9992 & 9.82 & 0.0040 & -1106.6 & 0.4005 & -605.3 \\
	$71s_{1/2}$ - $69s_{1/2}$ & $70p_{3/2}$ - $69p_{1/2}$ & 9.45 & 8.01 & 0.9987 & -10.66 & 0.0643 & -955.7 & 0.5348 & -518.5 \\
\hline\hline
    \end{tabular}}
    \caption{Rb-Cs \Forster resonances from Rb $s_{1/2}$ - Cs $s_{1/2}$ states previously identified \cite{beterov15}. $P_{1r}(t_\pi)$ and $P_{rr}(t_\pi)$ along with fitted $\tilde{C}_3$ and $\tilde{C}_6$ coefficients are calculated for $\theta=90^\circ$ and $(m_j^\mathrm{Rb},m_j^\mathrm{Cs})=(1/2,-1/2)$ assuming $\Omega/2\pi=2$~MHz.}    
    \label{tab:ss}
\end{table*}

\section{Two-Photon Excitation Parameters}
\label{sec:details}

In this paper we consider interactions between $d$-orbital Rydberg states of Cs and Rb which are created using a two-photon excitation process via an intermediate excited state. To include the effects of excitation and spontaneous decay from the intermediate states, we use the simplified effective Hamiltonian introduced in Eq.~\ref{eq:Heff} which is obtained by performing adiabatic elimination of excited state hyperfine levels $f_e$ requiring calculation of the following terms \cite{pelegri22}
\begin{eqnarray}
    \Omega_R =& \sum_{f_e} \frac{\Omega_1^{f_e}\Omega_{f_e}^r}{2\Delta_{f_e}}, \\
    \delta_\mathrm{AC} =& \sum_{f_e} \frac{\vert\Omega_1^{f_e}\vert^2-\vert \Omega_{f_e}^r\vert^2}{4\Delta_{f_e}}, \\ 
    P_e^1 =&\sum_{f_e} \frac{\vert\Omega_1^{f_e}\vert^2}{4\Delta^2_{f_e}},\\
    P_e^r =&\sum_{f_e} \frac{\vert\Omega^r_{f_e}\vert^2}{4\Delta^2_{f_e}},    
\end{eqnarray}
where $\Omega_R$ is the effective two-photon Rabi frequency, $\Delta_{f_e}$ is the single-photon detuning with respect to the intermediate hyperfine level, and $P_e^{1,r}$ is the residual excited state population due to coupling between the ground and Rydberg state transitions respectively.

Using this effective model to represent the transition from $\ket{1}$ and $\ket{r}$ to appear as a spin-$1/2$ system enforces the the desired result that after a $2\pi$ rotation the wavefunction acquires a $\pi$-phase shift. However for a real two-photon excitation with multiple intermediate excited states, the exact phase $\phi_{2\pi}$ accumulated after a $2\pi$-pulse can take any value between $0$--$2\pi$ dependent upon the relative power of the lasers and intermediate state detuning \cite{maller15}. To ensure accurate parameters are chosen for estimating the gate fidelity with the effective model, for each of the transitions in Cs and Rb we first generate an exact single atom model including all excited states without adiabatic ellimination, and adjust the power-ratios in the two lasers to tune the system to give an error in the phase-shift of $\vert \phi_{2\pi}-\pi\vert<10^{-5}$.

For both transitions we calculate single-photon Rabi frequencies $\Omega_1^{f_e}$ and $\Omega_{f_e}^r$ assuming a centre of mass detuning from the intermediate state of $10$~GHz, targetting a two-photon Rabi frequency of $\Omega_R/2\pi=5$~MHz typical of recent experimental demonstrations, and quote powers for a beam waist of 10~$\mu$m to allow easy scaling to other beam parameters.

Excitation in Rb from $\ket{1}=\ket{2,0}$ to $\ket{r}=\ket{59d_{5/2},m_j=5/2}$ with $\Gamma_r/2\pi=0.76$~kHz is modelled using transitions via the $6p_{3/2}$ intermediate state with $\Gamma_e/2\pi=1.4$~MHz using a pair of $\sigma^+$-polarized photons. Choosing powers of 20~mW and 150~mW for the lower and upper transitions results in a $\pi$-phase shift with $P_e^{1,r}=(5.0,1.7)\times10^{-4}$ respectively. Excitation in Cs from $\ket{1}=\ket{4,0}$ to $\ket{r}=\ket{68d_{3/2},m_j=-3/2}$ with $\Gamma_r/2\pi=0.86$~kHz is modelled using transitions via the $7p_{1/2}$ intermediate state with $\Gamma_e/2\pi=1.0$~MHz using a pair of $\sigma^-$-polarized photons. Choosing powers of 5.2~mW and 240~mW for the lower and upper transitions results in a $\pi$-phase shift with $P_e^{1,r}=(5.0,1.2)\times10^{-4}$ respectively.

\end{document}